\documentclass[10pt,conf]{IEEEtran}
\ifCLASSINFOpdf \else \fi
\usepackage{geometry}
\usepackage{multirow}
\usepackage{bbold}
\usepackage{ifpdf}
\usepackage[normalem]{ulem} 
\usepackage{subcaption}%
\usepackage[cmex10]{amsmath}
\usepackage{amssymb}
\usepackage{verbatim}
\usepackage{algorithm}
\usepackage{array}
\usepackage{latexsym}
\usepackage{color}
\usepackage{textgreek}
\usepackage{subscript}
\usepackage{rotating}
\usepackage{booktabs,multirow}
\usepackage{mdframed}	
\usepackage{tabularx}
\usepackage{amsmath}
\usepackage{makecell}
\usepackage{tabto}
\usepackage{mathrsfs}
\usepackage{url}
\usepackage{cite}
\usepackage{listings}
\usepackage{array}
\usepackage[table]{xcolor}
\usepackage{ragged2e}
\usepackage{booktabs}

\usepackage[noend]{algpseudocode}
\usepackage{algorithm}

\usepackage{graphicx}
\usepackage{comment}
\usepackage[cmex10]{amsmath}
\usepackage{amssymb}
\usepackage{bbold}
\usepackage{float}
\usepackage[cmex10]{amsmath}
\usepackage{amssymb}
\usepackage{verbatim}
\usepackage{array}
\usepackage{latexsym}
\usepackage{color}
\usepackage{tabularx} 
\usepackage{textgreek}
\usepackage{subscript}
\usepackage{rotating}
\usepackage{booktabs,multirow}
\usepackage{mdframed}	
\usepackage{tabularx}
\usepackage{amsmath}
\usepackage{makecell}
\usepackage{tabto}
\usepackage{mathrsfs}
\usepackage{url}
\usepackage{cite}
\usepackage{listings}
\usepackage{array}
\usepackage[table]{xcolor}

\newcommand{\notered}[1]{\textcolor{red}{[{\bf #1}]}}

\definecolor{pinkpurple}{rgb}{0.6, 0.1, 0.9} 
\usepackage{array}
\usepackage{soul}
\sethlcolor{green}
\usepackage{url}
\usepackage{pifont}
\usepackage{xcolor}
\usepackage{multirow}
\usepackage{graphicx} 

\begin{document}

\title{ Efficient Adversarial Detection Frameworks for Vehicle-to-Microgrid Services in Edge Computing}

\author{
	Ahmed Omara~\IEEEmembership{Member~IEEE},~and~Burak Kantarci,~\IEEEmembership{Senior Member,~IEEE}
\thanks{
The authors are with the School of Electrical Engineering and Computer Science at the University of Ottawa, Ottawa, ON, K1N 6N5, Canada.
E-mail: \{aomar020,burak.kantarci\}@uottawa.ca}  
}

\maketitle
\thispagestyle{empty}
\pagestyle{empty}
\begin{abstract}
As Artificial Intelligence (AI) becomes increasingly integrated into microgrid control systems, the risk of malicious actors exploiting vulnerabilities in Machine Learning (ML) algorithms to disrupt power generation and distribution grows. Detection models to identify adversarial attacks need to meet the constraints of edge environments, where computational power and memory are often limited. To address this issue, we propose a novel strategy that optimizes detection models for Vehicle-to-Microgrid (V2M) edge environments without compromising performance against inference and evasion attacks. Our approach integrates model design and compression into a unified process and results in a highly compact detection model that maintains high accuracy. We evaluated our method against four benchmark evasion attacks-Fast Gradient Sign Method (FGSM), Basic Iterative Method (BIM), Carlini \& Wagner method (C\&W) and Conditional Generative Adversarial Network (CGAN) method—and two knowledge-based attacks, white-box and gray-box. Our optimized model reduces memory usage from 20MB to 1.3MB, inference time from 3.2 seconds to 0.9 seconds, and GPU utilization from 5\% to 2.68\%.\end{abstract}

\begin{IEEEkeywords}
vehicle-to-microgrid (V2M), machine learning, smart microgrids, cybersecurity, generative adversarial network (GAN), evasion attack, inference attack,  vehicle-to-home (V2H)
\end{IEEEkeywords}

%
\IEEEpeerreviewmaketitle

\section{Introduction}
\label{sec:intro}
V2M systems represent a pivotal component of transactive energy frameworks, leveraging Electric Vehicles (EVs) as decentralized power sources to supply excess electricity to local smart microgrids~\cite{simsek2020cost}. These systems become particularly critical during power outages, enabling direct energy transactions between EVs and smart grids. This framework highlights the potential of EVs to enhance the stability and reliability of smart grids by establishing direct connections that surpass traditional grid dependencies~\cite{omara2021impact}. V2M systems use smart meters to monitor and report the energy consumption of each microgrid to electric power companies using wireless communication technologies~\cite{icc2018}. Despite their utility, smart meters are susceptible to adversarial data attacks that manipulate electricity measurements, consequently disrupting the dispatch of EV resources to affected microgrids. To address such threats, ML models have been employed as a means of detecting anomalies and attacks. However, these models themselves face vulnerabilities to adversarial machine learning attacks, where adversaries carefully craft perturbations that mislead ML models into incorrect predictions~\cite{OMARA2024103016}. For example, in detecting data integrity attacks against smart meters, adversaries might execute an evasion attack to deceive the ML classifier. Such an attack can cause a maliciously altered electricity reading to be mistakenly classified as benign. This targeted ML model, henceforth referred to as the victim’s ML classifier, becomes a critical focal point in safeguarding the integrity of V2M services.\\
 Traditional methods of processing the vast amount of data transmitted from smart meters to centralized cloud servers is becoming less feasible due to latency and bandwidth issues, necessitating real-time, localized processing. This need has led to a shift towards Edge AI, where ML algorithms are deployed directly on edge devices like IoT sensors and mobile phones, enabling data to be processed locally~\cite{sipola2022artificial}. This local processing reduces latency, enhances privacy and security, and decreases network bandwidth use. However, deploying ML on edge devices poses challenges due to their limited processing power, memory, and network bandwidth, along with constraints on battery life and power consumption~\cite{kum2022optimization}. Addressing these challenges requires model compression techniques to adapt ML models to the limited capabilities of edge devices. Optimizing ML models for edge deployment allow smart meters in smart grids to process data more efficiently, enhancing their functionality and sustainability~\cite{chang2021survey}. \\
Most existing studies, such as~\cite{takiddin2022robust} and~\cite{l14}, focus on defending against adversarial attacks without considering model efficiency, making them unsuitable for edge deployment. Other works, such as~\cite{zhang2015efficient} and~\cite{surianarayanan2023survey}, apply compression techniques but do not evaluate adversarial robustness. Our work bridges this gap by integrating adversarial robustness and efficient model compression in V2M settings, ensuring both high detection accuracy and resource efficiency for real-time edge applications. Furthermore, our approach significantly reduces the computational footprint while maintaining robust adversarial detection, making it a practical solution for securing smart microgrids in constrained environments. In our recent work~\cite{omara2022Globecom}, we have demonstrated how susceptible V2M systems are to adversarial attacks, especially when adversaries use the Conditional Generative Adversarial Network (CGAN) model to increase the impact and success rate of their attacks. In this work, we implement a compact detection model by implementing an algorithm for model design and model compression. To the best of our knowledge, for the first time, this work proposes an end-to-end model optimization for adversarial detection models under V2M settings. The main contributions of our work are as follows:
\begin{itemize}
\item We propose an end-to-end optimized detection framework for V2M systems under adversarial attacks, integrating NAS-based model selection and State-of-The-Art (SoTA) compression techniques such as pruning, quantization, projection. Our method significantly reduces memory footprint (from 20MB to 1.3MB), inference time (from 3.2 to 0.9 s) GPU utilization (from 5\% to 2.68\%) while maintaining high adversarial detection accuracy.
\item  We evaluate five access cases to the victim's ML classifier training dataset, with four cases of gray-box attacks and one case of white-box attack. In the gray-box attack, adversaries can have different access percentages to the victim’s ML training dataset, namely 20\%, 40\%, 60\% or 80\%. In the white-box attack, adversaries can have access to 100\% of the victim’s ML training dataset. 
\item Study the impact of the types of benchmark evasion attack, namely FGSM~\cite{goodfellow2014explaining}, BIM~\cite{kurakin2018adversarial}, C\&W~\cite{carlini2017towards} and CGAN method~\cite{omaravcc}. We show that our optimized detection model can effectively detect different evasion attacks while reducing the model's size and inference time.
\item Present a compressed Convolution Neural Network (CNN) to prevent adversarial attacks that are augmented using a CGAN model as studied in~\cite{OMARA2024103016}. We show that our compact detector can achieve high detection performance. 
\end{itemize}

\section{Background and Related Work} \label{sec:optimization}
Deep learning models commonly use 32-bit floating-point parameters. Model quantization compresses these models by reducing the bit precision of their parameters, thereby decreasing memory requirements and potentially speeding up inference. While 8-bit quantization has been studied and delivers strong performance~\cite{wang2022learnable}, more extreme approaches—such as 1-bit quantization have also proven effective~\cite{gholami2022survey}. Notably, these lower bit-precision models can still achieve near full-precision accuracy, particularly in larger and over-parameterized networks~\cite{krishnamoorthi2018quantizing}. Quantized models retain the same network structure as their full-precision versions, hence, developing and deploying quantization schemes is significantly more straightforward than other compression strategies. 
Processors can handle operations involving 8-bit integers faster than those involving 32-bit floating points due to the reduced data width, which accelerates the overall processing time of the neural network~\cite{krishnamoorthi2018quantizing}.

Similar to quantization, pruning techniques have also been used to compress deep learning models. Over-parameterized networks during training provides the flexibility needed for effective learning, even though smaller models may struggle to capture the data’s underlying patterns. After training, however, many connections prove to be redundant or irrelevant, allowing pruning methods to remove them with little impact on performance. This process yields sparser models that maintain comparable accuracy. The authors in~\cite{frankle2018lottery} proposed viewing pruning as a way to uncover a subnetwork within an over-parameterized model. Because many parameters are redundant, the pruned subnetwork can be retrained from scratch using the original random initialization, bypassing the overhead of training unnecessary parameters. This strategy is conceptually similar to the teacher assistant knowledge distillation method in~\cite{mirzadeh2020improved} and involves iteratively applying structured single-shot pruning until the target level of sparsity is reached.

In~\cite{zhang2015efficient}, the authors implemented a projection-based method that uses Principal Component Analysis (PCA) for compression. The method capitalizes on PCA to efficiently reduce the dimensionality of the layers' responses (i.e., output), which are inherently high-dimensional and computationally expensive to process. To improve the accuracy of the projection-based method, the authors in~\cite{yu20182dpcanet} used 2D PCA for filter analysis, giving rise to the 2D PCA-Net method.

\begin{figure*}[t]
  \centering
\includegraphics[height=3.25cm,width=15.cm,scale=1]{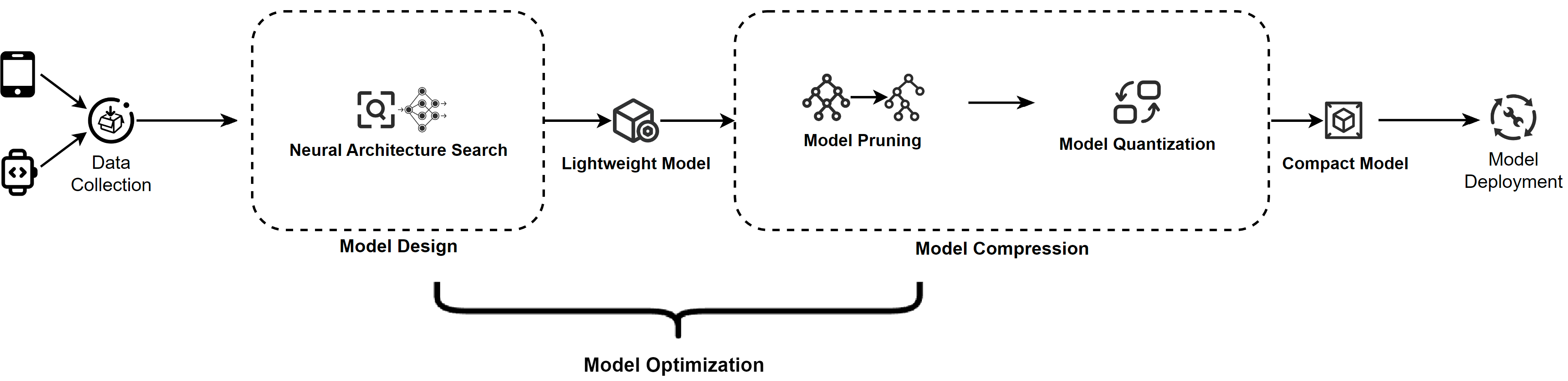}
  \caption{AI deployment pipeline}
  \label{modeldeployment}
\end{figure*} 

Machine learning-based attack detection in V2M systems are often computationally expensive, making them unsuitable for edge deployment. Existing works either focus solely on improving adversarial robustness or optimizing model efficiency, but not both simultaneously. To address this, we propose a novel e2e optimized adversarial detection framework that balances robustness and efficiency in V2M edge environments.

\section{Proposed Framework}~\label{framework}
In Fig.~\ref{modeldeployment}, we view the end-to-end pipeline for detection model deployment. As part of this process, model optimization is crucial for creating a compact and robust model in edge environments. A critical aspect of model optimization is model design, which is an exhaustive stage that requires extensive domain knowledge to select the right model architecture and hyperparameters. For this part, we use Neural Architecture Search (NAS) to find the best model architectures that deliver the best performance in terms of accuracy and F1-score. We then implement model compression techniques to reduce the model's size while maintaining optimal performance.

\begin{algorithm}
\caption{NAS with Bayesian Optimization Under Specific Performance Constraints}
\label{algorithm2}
\begin{algorithmic}[1]
\State \textbf{Input:} Training data \( D_{\text{train}} \), validation data \( D_{\text{val}} \), accuracy threshold \( C \), F1-score threshold \( F \), number of top architectures \( n \)
\State \textbf{Output:} Set of top \( n \) best performance CNN architectures \( \{M_i\} \)

\Function{NAS\_Bayesian\_Optimal}{$D_{\text{train}}, D_{\text{val}}, C, F$}
    \State Define search space \( S \)
    \State Initialize Bayesian model \( B \)
    \State Define acquisition function \( \alpha \)
    \While{not stopping\_criterion}
        \State \( A \gets \) select\_next\_architecture(\( B, \alpha \))
        \State \( \text{perf} \gets \) train\_and\_evaluate(\( A, D_{\text{train}}, D_{\text{val}} \))
        \State update\_Bayesian\_model(\( B, A, \text{perf} \))
    \EndWhile
    \State \( \text{Top}_n \gets \) select\_top\_n\_architectures(\( B, n, C, F \))
    \State \Return \( \text{Top}_n \)
\EndFunction

\State \( \text{Top}_n \gets \) \Call{NAS\_Bayesian\_Optimal}{$D_{\text{train}}, D_{\text{val}}, C, F$}
\end{algorithmic}
\end{algorithm}

We explain our methodology here in details. In Algorithm~\ref{algorithm2}, we implement NAS method using Bayesian Optimization~\cite{white2021bananas}, which will identify best performing CNN architectures regardless of their size since the model compression will be part of the next algorithm. This process begins by defining the search space \(S\), which encompasses various possible configurations of CNN architectures, including different types and arrangements of layers, activation functions, and other hyperparameters. The core of the NAS process involves a Bayesian optimization framework, initialized by setting up a Bayesian model \(B\). This model is key in predicting the performance of various architectural configurations based on historical evaluation data. The algorithm employs an acquisition function \( \alpha \), which guides the search. It helps in balancing the exploration of new, untested architectural configurations against the exploitation of configurations known to perform well, thus ensuring an efficient search process.\\
During the NAS execution, the algorithm iteratively selects the next architecture to evaluate by utilizing the acquisition function. Each selected architecture is then trained and assessed on the given training and validation datasets \(D_{\text{train}}\) and \(D_{\text{val}}\), respectively. The performance of each architecture is measured, and the results are used to update the Bayesian model, enhancing its prediction accuracy for future iterations. This loop continues until the predefined stopping criteria \(C\) and \(F\) are met. The stopping criteria serve different purposes. For instance, it acts as a guide for the Bayesian optimization process to explore the search space efficiently. It does not only facilitate a more structured search but also aid in evaluating the success of the search strategy itself. Without such criteria, the optimization process will not have metrics to assess the performance of CNN architectures. This will lead to continuously explore the search space and without exploiting. Although the performance metrics can be set to (1.0) (as we did in this work), in practical scenarios, especially in constrained environment, achieving accuracy and F1-score of (1.0) may not be feasible due to limitations in computational resources and memory. Moreover, depending on the complexity of the application and dataset, achieving accuracy of (1.0) could be a sign of overfitting~\cite{cogswell2015reducing}. Hence, users can set lower performance criteria to avoid such issues. Besides, different applications may have varying requirements for accuracy and F1-score based on their specific needs and the consequences of misclassifications. Once the search is complete, the algorithm selects the top \(n\) architectures that not only meet but potentially exceed the defined accuracy and F1-score thresholds. This selection is based purely on performance metrics, without considering the computational or memory efficiency of the architectures. These top-performing models form the output of the NAS process and are poised for subsequent optimization steps, such as pruning and quantization, to enhance their efficiency for deployment in edge environments.\\
\begin{algorithm}
\caption{Best Optimized and Compressed Architecture}
\label{algorithm3}
\begin{algorithmic}[1]

\State \textbf{Input:} Set of top \( n \) optimized CNN architectures \( \{M_i\} \), Training data \( D_{\text{train}} \), Validation data \( D_{\text{val}} \)
\State \textbf{Output:} Optimized and compressed CNN model \( M_{\text{best}} \)

\Function{Prune\_Quantize\_Models}{$\{M_i\}, D_{\text{train}}, D_{\text{val}}$}
    \State Initialize \( \text{BestScore} \gets -\infty \)  
    \State Initialize \( M_{\text{best}} \)
    \For{\( M \) in \( \{M_i\} \)}
        \State \( M_{\text{trained}} \gets \) train\_model(\( M, D_{\text{train}} \))
        \State \( M_{\text{pruned}} \gets \) iterative\_pruning(\( M_{\text{trained}}, D_{\text{train}} \))
        \State \( M_{\text{quant}} \gets \) quantize\_model(\( M_{\text{pruned}}, D_{\text{train}}, D_{\text{val}} \))
        \State \( M_{\text{quant}} \gets \) fine\_tune(\( M_{\text{quant}}, D_{\text{train}}, D_{\text{val}} \))
        \State \( \text{Accuracy, F1Score} \gets\) evaluate\_Accuracy\_F1Score (\( M_{\text{quant}}, D_{\text{val}} \))
        \State \( \text{ModelSize} \gets \) get\_model\_size (\( M_{\text{quant}} \))
        \State \( \text{Score} \gets \) calculate\_CompositeScore (\text{Accuracy}, \text{F1Score}, \text{ModelSize})
        \If{\( \text{Score} > \text{BestScore} \)}
            \State \( \text{BestScore} \gets \text{CompositeScore} \)
            \State \( M_{\text{best}} \gets M_{\text{quant}} \)
        \EndIf
    \EndFor
    \State \Return \( M_{\text{best}} \)
\EndFunction

\State \( M_{\text{best}} \gets \) \Call{Prune\_Quantize\_Models}{$\{M_i\}, D_{\text{train}}, D_{\text{val}}$}
\end{algorithmic}
\end{algorithm}
Following the selection of top-performing architectures algorithm.1, the next phase involves refining these models through pruning and quantization to enhance their computational efficiency. 
Algorithm~\ref{algorithm3} begins by taking each architecture from the set of top \(n\) optimized CNN architectures. Initially, every model is trained to establish a strong baseline before any modifications that could potentially degrade the model's effectiveness. Once the models are fully trained, the next step is iterative pruning. During this stage, the algorithm systematically removes the least important connections within each model. Taylor-based pruning is used in this step. This method effectively reduces the model's complexity and size, which in turn decreases memory usage and computational demands without significantly sacrificing accuracy.\\
Following pruning, the models are subjected to quantization. This process converts the model's floating-point weights to a lower-precision format, such as 8-bit integers. Quantization significantly reduces the memory footprint of the model and is often accompanied by speed improvements during inference. However, quantization can sometimes lead to a loss in accuracy, which is why the next crucial step is fine-tuning. Fine-tuning the quantized models using both training and validation data helps recover any accuracy lost during the quantization process. It adjusts the model parameters within the constraints of their new precision levels to ensure the final model continues to perform robustly. \\
Subsequently, the algorithm evaluates each model on three parameters: accuracy, F1-score, and the size of the model. Accuracy and F1-score are direct indicators of the model’s performance, assessing how well the model predicts correct outcomes and balances precision and recall, respectively. The size of the model is also evaluated to ensure the final model is not only high-performing but also compact enough for efficient deployment. To synthesize these diverse metrics into a singular decision metric, the algorithm computes a composite score for each model. This composite score is calculated through a weighted sum function as described in equation~\ref{ee3}. For simplicity, we use equal weights for all three metrics.
\begin{eqnarray}
 Composite Score = w_{acc} \cdot Accuracy + w_{F1} \cdot \text{F1-score} +\nonumber\\ w_{size} \cdot \left(1 - \frac{\text{Compressed Size}}{\text{Original Size}}\right)
   \label{ee3}    
 \end{eqnarray} 




\section{Performance Evaluation} \label{sec:performance}

\subsection{Simulation settings}
The CNN network comprises of multiple convolutional layers. The first layer uses 128 filters of size 3 with a stride of 1, employing the ReLU activation function. This is followed by
two convolutional layers with 256 filters of the same size and stride to enhance feature extraction capabilities. The MaxPooling layer with a pool size of 2 is applied next. The final convolutional layer employs 512 filters to capture more complex patterns before the data is flattened into a one-dimensional array to facilitate dense layer processing. The dense segment of the network features a fully connected layer with 1024 units using ReLU activation, designed to synthesize the features extracted in previous layers. A dropout rate of 50\% is implemented to prevent overfitting. The output layer is configured for a binary classification task which uses a sigmoid activation function to produce a probability indicating if the input being benign or malicious input. In addition, we compare five cases where the adversary has access to varying amounts of the victim's ML training dataset. In white-box attacks, the adversary has access to 100\% to the training dataset. Gray-box attacks, the adversary has 80\%, 60\%, 40\% and 20\% access to the training dataset. In consideration of the involved components, the following two assumptions are made:
(i) Adversaries in V2M settings can partially/fully access a read-only version of the training dataset of victim's ML model~\cite{energydatabreaches2023}. Alternatively, adversaries can build a shadow training dataset by querying the victim's model and observe the input-output data~\cite{chakraborty2021survey}.
(ii) Adversaries can perform False Data Injection Attack (FDIA) by exploiting the smart meters vulnerabilities as in~\cite{3}\cite{8}.\\
We use the EMSx dataset~\cite{le2023emsx}, provided by Schneider Electric, which is a comprehensive collection of data tailored for the analysis and management of electric microgrids. It encompasses historical observations and forecasts related to photovoltaic generation and energy demand across 70 industrial sites. This dataset is particularly designed for developing and testing control algorithms for microgrids that include photovoltaic units and energy storage systems. Each site in the dataset is well-documented with specific parameters related to battery operation and time-series data on energy usage and production. The goal of the adversary is to create stealthy perturbations that will be introduced to smart meter readings through the meters vulnerabilities to compromise the victim's ML classifier. We chose the EMSx dataset due to its comprehensive representation of real-world industrial microgrid scenarios. We plan to validate our framework on additional datasets to assess cross-domain robustness.\\
To simulate adversarial activities, we constructed a malicious dataset using several functions designed to emulate various realistic malicious behaviors, as in~\cite{jokar2015electricity}, as well as adversarial attacks generated by evasion methods. These include partial reductions in reported power (f1(·) and f2(·)), selective bypassing of reporting (f3(·)), and manipulations based on time-of-use (ToU) pricing (f4(·), f5(·), and f6(·)). Moreover, we evaluate the ADR performance under three evasion techniques, namely FGSM, BIM and C\&W. We provide a short overview of each technique as below.
\begin{enumerate}
    \item FGSM technique: This method adds perturbation to input ($x$) in the direction of the sign of the gradient of the cost function ($J$) to create a new ($x'$) that maximizes the loss. The perturbation's magnitude is regulated by the parameter $\epsilon$, defined as in Eq.~\ref{eq5}.
    \begin{eqnarray}
    x' = x + \epsilon \cdot \text{sign}(\nabla_x J(\theta, x, y))
    \label{eq5}
    \end{eqnarray}
    \item BIM technique: This attack iteratively applies FGSM in small steps \(\alpha\), clipping intermediate values to stay within an \(\epsilon\)-neighborhood of the original input ($x$). The process is defined as in Eq.~\ref{eq6}.
    \begin{eqnarray}
    x'_{n+1} = \text{Clip}_{x, \epsilon} \{ x'_n + \alpha \cdot \text{sign}(\nabla_x J(\theta, x'_n, y)) \}
    \label{eq6}
    \end{eqnarray}
    \item C\&W technique: This complex attack optimizes to find the smallest perturbation \(\delta\) causing misclassification. It uses a regularization parameter ($c$) determined by binary search. The goal is to minimize the objective function \(f\), achieving its lowest value when the output nears the decision boundary, as shown in Eq.~\ref{eq7}
    \begin{eqnarray}
    \text{Minimize } \| \delta \|_2 + c \cdot f(x + \delta) 
    \label{eq7}
    \end{eqnarray}
\end{enumerate}
\subsection{Numerical Results}
\begin{figure}[H]
 \centering
\includegraphics[height=5.6cm,width=8.75cm,scale=0.25]{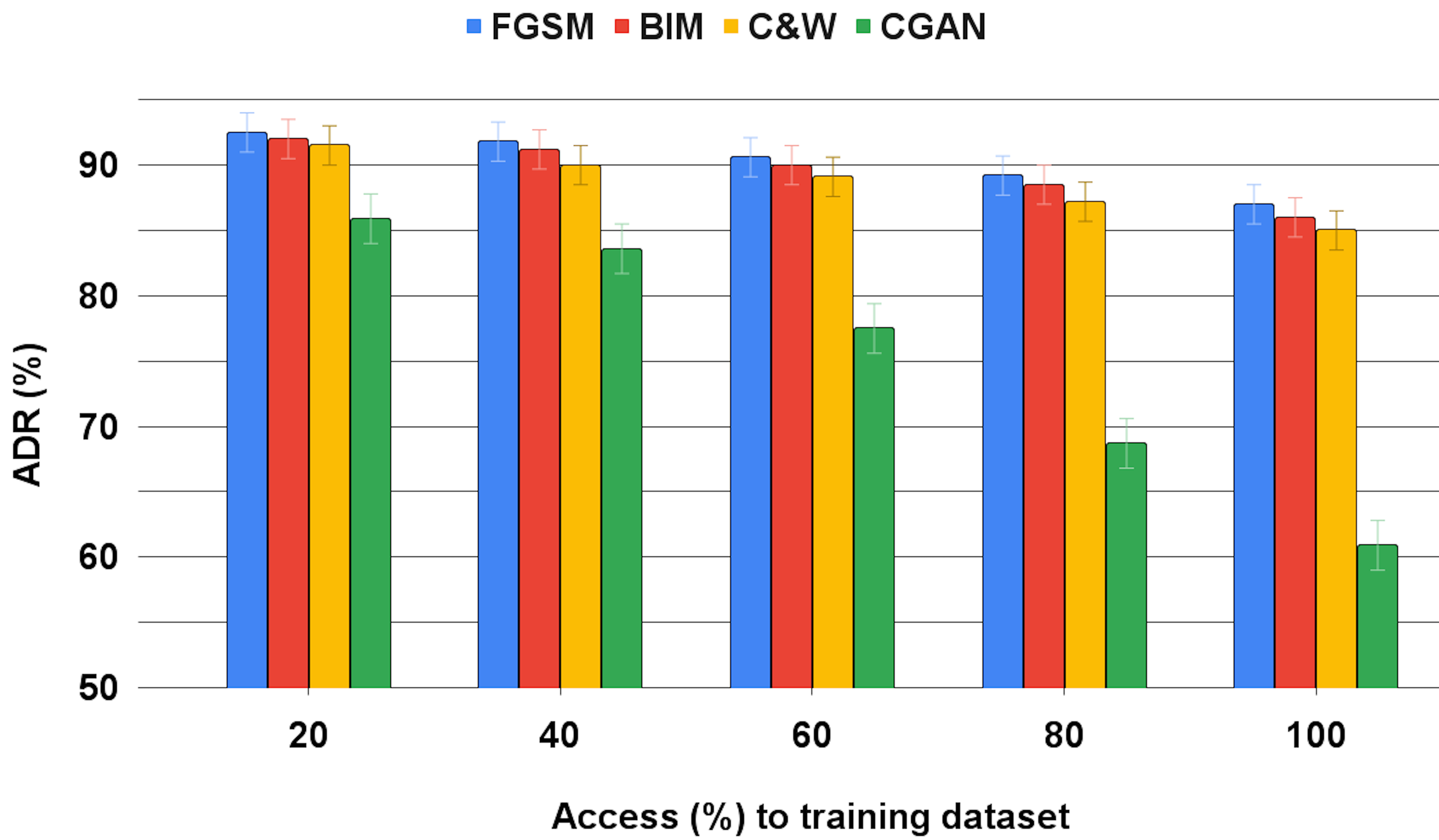}
  \caption{CNN detection model performance before compression }
  \label{ganbeforecompression}
\end{figure}

The GAN-based detection method proves to be more effective in identifying and detecting adversarial attacks as shown in~\cite{10475111}. This is due to the ability of generative models (i.e., GAN) to capture the underlying distribution of legitimate data more accurately, enabling them to better model deviations caused by adversarial perturbations. In this work, we only focus on optimizing the detection model.
\begin{table}
\centering
\caption{Proposed method vs. SoTA methods}
\label{tab:model_performance}
\begin{tabular}{|p{0.5in}|p{0.4in}|p{0.4in}|p{0.5in}|p{0.4in}|}
\hline
\textbf{Model Type} & \textbf{Inference Time (s)} & \textbf{Memory Space (MB)} & \textbf{GPU Utilization (\%)} & \textbf{Accuracy (\%)} \\ \hline
Original Model & 3.2 & 20 & 5\% & 94.5\% \\ \hline
Quantized Model & 2.74 & 6.5 & 3.75\% & 89.9\% \\ \hline
Pruned Model & 2.4 & 13 & 4.62\% & 87\% \\ \hline
Projected Model & 2.1 & 9.5 & 4.30\% & 88.5\% \\ \hline
\textbf{Proposed Method} & \textbf{0.9} & \textbf{1.35} & \textbf{2.68\%} & \textbf{92.7\%} \\ \hline
\end{tabular}
\end{table}
Table~\ref{tab:model_performance} provides a detailed comparison of various models in terms of inference time, memory space, GPU utilization and accuracy. Notably, the proposed method significantly outperforms the other models across all metrics. With an inference time of just 0.9 seconds, the proposed method is markedly faster than the original model, which takes 3.2 seconds. This substantial improvement in inference time reduces the processing time by over 70\%, hence, enhancing the efficiency of real-time applications. In terms of memory space, the proposed method also excels, requiring only 1.35 MB, when compared to the original model’s 20 MB. This substantial decrease in memory usage highlights the method’s efficiency and suitability for constrained environments. Furthermore, the GPU utilization of the proposed method is only 2.68\%, significantly lower than the original model’s 5\%. This reduction in utilization not only conserves computational resources but also reduces power consumption, making the proposed method more environmentally friendly and cost-effective. Although model compression typically results in reduced accuracy, the proposed method outperforms other compression techniques, achieving an accuracy of 92.7\%. Additionally, the proposed method shows improvements over the quantized, pruned, and projected models in all aspects, demonstrating its robustness and superior design. These results indicate that the proposed method not only enhances performance but also optimizes resource usage. The GAN-based detection system stands out for its resilience, maintaining relatively high detection rates even at full dataset access as shown in Fig.~\ref{ganbeforecompression}. The detector's performance against FGSM attacks dips from 92.5\% to 87\% as dataset access grows from 20\% to 100\%. 
\vspace{-0.35cm}
\vspace{-0.25cm}
\begin{figure}[H]
  \centering
\includegraphics[height=5.6cm,width=8.75cm,scale=0.25]{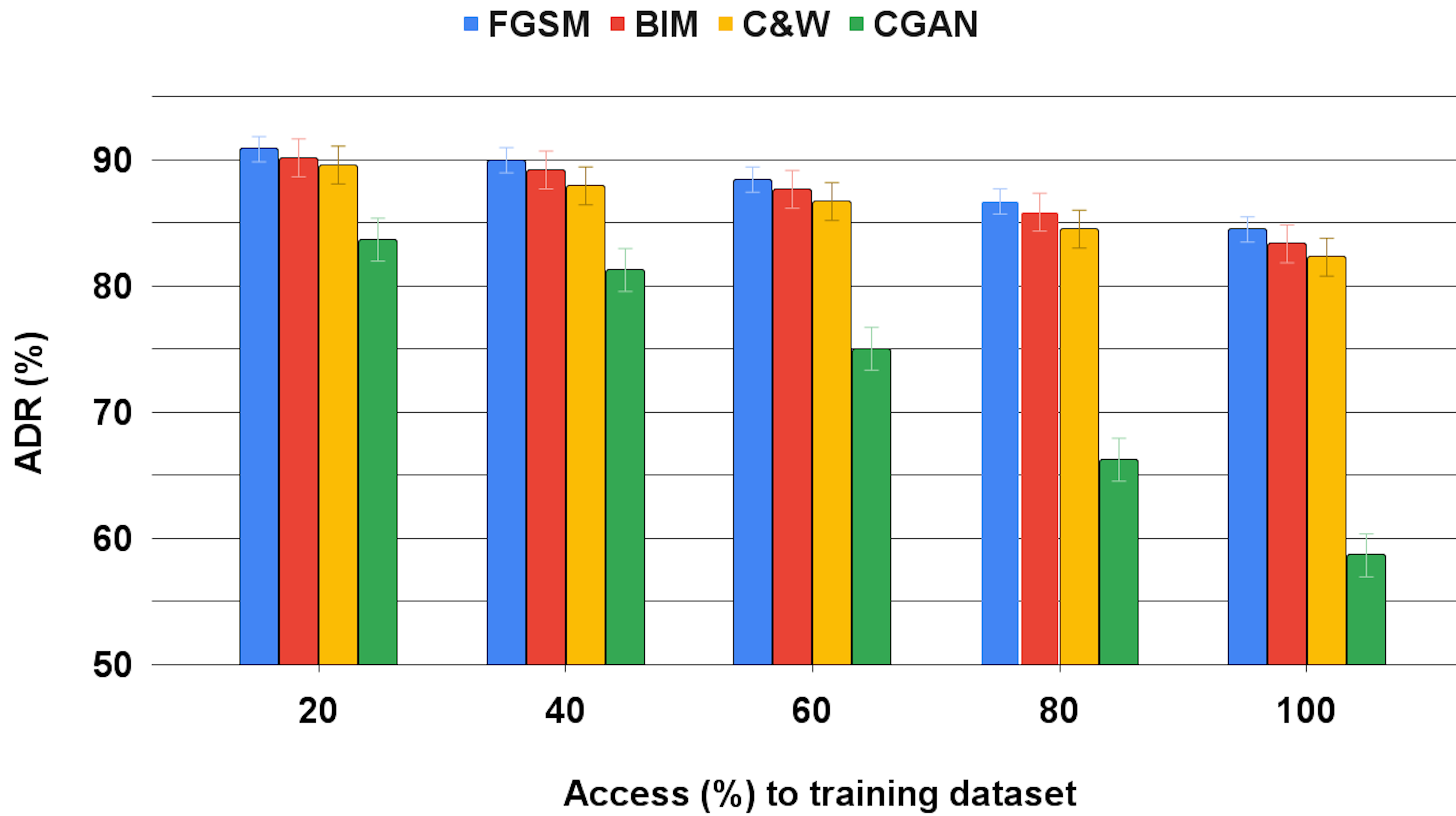}
\caption{CNN detection model performance after compression}
  \label{ganaftercompression}
\end{figure}

Performance metrics for the CNN detector under various evasion methods and differing levels of access to the training dataset reveal significant trends both before and after the application of a compression method. Before compression, the detection performance generally shows a decline as the adversary's access to the training dataset increases, as depicted in Fig.~\ref{ganbeforecompression}. For FGSM, the detection rate starts at 92.5\% when the adversary has 20\% access and drops to 87\% at 100\% access. A similar trend is observed for BIM, starting at 92\% and decreasing to 86\%. The C\&W attack sees a slight drop from 91.5\% to 85\%. The most significant drop is seen with CGAN, where detection performance decreases from 85.9\% to 60.9\%. After the compression method, a minimal reduction in detection performance across all attack methods occurs (Fig.~\ref{ganaftercompression}). For FGSM, the detection rate decreases from 91\% at 20\% access to 84\% at 100\% access, showing a similar trend to the pre-compression results but at slightly lower detection rates. For BIM, the detection rate drops from 90.2\% to 83.3\%, and for C\&W, it decreases from 89.6\% to 82.3\%. 
\section{Conclusion}
\label{sec:conclusion}
In this paper, we have addressed the challenge of adapting adversarial detection models for edge devices in V2M environments with limited computational power and memory.
 Focusing on reducing model size without compromising performance, we have explored AI model compression techniques such as projection, pruning, and quantization. Our proposed method integrates NAS with Bayesian Optimization offline as a one-time model design step, and not during real-time operation. This allows us to generate of compact and high-performing models tailored for different resource configurations. Furthermore, as a first step toward real-world deployment, we focused on rigorous simulation-based evaluation using realistic adversarial attack scenarios and performance constraints relevant to edge devices. While we did not conduct physical deployment in this work, our optimization results—achieving 92.7\% accuracy with only 1.35 MB model size and 0.9s inference time—demonstrate practical viability for edge implementation. 
\vspace{-0.1cm}
\section*{Acknowledgements}
This work is supported in part by National Science and Engineering Research Council (NSERC) Discovery and NSERC CREATE TRAVERSAL programs.
\bibliographystyle{IEEEtran}

\end{document}